\documentclass[prd,twocolumn]{revtex4}
\usepackage{amsmath, graphicx}
\usepackage{grffile}
\usepackage{dcolumn}
\usepackage{bm}
\usepackage{epsfig}
\usepackage{mathrsfs}  
\usepackage{subfigure}
\usepackage{multirow}
\usepackage{epstopdf}
\usepackage{amsmath}
\usepackage{algorithmicx}
\usepackage{amssymb}
\usepackage{tensor}
\usepackage[math]{cellspace}
\usepackage{bookmark}
\usepackage[usenames,dvipsnames]{xcolor}
\usepackage{hyperref}
\usepackage{slashed}
\usepackage{youngtab}
\usepackage{physics}
\usepackage{tensor}
\usepackage[normalem]{ulem}

\makeatletter
\renewcommand*\env@matrix[1][\arraystretch]{%
  \edef\arraystretch{#1}%
  \hskip -\arraycolsep
  \let\@ifnextchar\new@ifnextchar
  \array{*\c@MaxMatrixCols c}}
\makeatother

\begin{document}

\title{Nonlinear Dynamics of Quadratic Gravity in Spherical Symmetry}

\author{Aaron Held}%
 \email{a.held@imperial.ac.uk}
\affiliation{Blackett Laboratory, Imperial College London, London SW7 2AZ, United Kingdom}

\author{Hyun Lim}
\email{hyunlim@lanl.gov} 
\thanks{\\The two authors contributed equally. The order of authors is listed alphabetically.}
 \affiliation{Computational Physics and Methods (CCS-2), Los Alamos National Laboratory, Los Alamos, NM 87545 USA}
 \affiliation{Center for Theoretical Astrophysics, Los Alamos National Laboratory, Los Alamos, NM 87545 USA}
\affiliation{Applied Computer Science (CCS-7), Los Alamos National Laboratory, Los Alamos, NM 87545 USA}
\date{\today}

\begin{abstract}
We present the first numerically stable nonlinear evolution for the leading-order gravitational effective field theory (Quadratic Gravity) in the spherically-symmetric sector.
The formulation relies on (i) harmonic gauge to cast the evolution system into quasi-linear form (ii) the Cartoon method to reduce to spherical symmetry in keeping with harmonic gauge, and (iii) order-reduction to 1st-order (in time) by means of introducing auxiliary variables.
Well-posedness of the respective initial-value problem is numerically confirmed by evolving randomly perturbed flat-space and black-hole initial data. Our study serves as a proof-of-principle for the possibility of stable numerical evolution in the presence of higher derivatives.
\end{abstract}
\maketitle

\section{Introduction}
\label{sec:intro}

General Relativity (GR) is in excellent agreement with an ever-growing body of experimental tests. At the same time, theoretical considerations strongly suggest that the theory is incomplete:
It is not known how to consistently couple GR to quantum field theories of matter once back-reactions are non-negligible.
Moreover, even classical matter distributions are prone to gravitational collapse and thus to the formation of curvature singularities, typically accompanied by geodesic incompleteness. See~\cite{Penrose:1964wq, Hawking:1965mf, Geroch:1966ur, Hawking:1969sw} for singularity theorems in static and highly symmetric settings and, e.g.,~\cite{Dafermos:2003wr, Harada:2001nj}
for numerical explorations in less symmetric and dynamical settings. 

Said theoretical breakdowns strongly motivate to embed GR into the modern framework of effective field theory (EFT) and treat the Einstein-Hilbert action merely as the leading-order term in a local-curvature expansion of a general diffeomorphism-invariant action of gravitational (and matter) degrees of freedom. Inconsistencies in the coupling to matter and the formation of singularities can then be interpreted as a consequence of extrapolating the EFT beyond its regime of validity: As the curvature grows during gravitational collapse, higher-order terms in the EFT will eventually become non-negligible and alter the dynamics of GR at some, as of now untested, curvature scale. We may encounter this scale anywhere between the largest currently probed curvature scales and the Planck scale. 
Indeed, quantum fluctuations are widely expected to induce such EFT curvature corrections, cf.~\cite{tHooft:1974toh,Stelle:1976gc,Goroff:1985sz,Avramidi:1985ki,vandeVen:1991gw} for perturbative quantum gravity, \cite{Boulware:1985wk, Zwiebach:1985uq}~for string theory, and~\cite{Percacci:2017fkn,Eichhorn:2018yfc,Reuter:2019byg} for reviews in the context of asymptotic safety.

Since GR tends to hide regimes of growing curvature behind event horizons~\cite{Penrose:1969pc}, experimentally probing the horizon-scale physics of black holes is presumably one of the most promising ways to push the limits of the EFT of gravity. A rapidly increasing number of gravitational-wave (GW) events from black-hole binary mergers~\cite{TheLIGOScientific:2016src,GWTC2Ligo2020} provide access to this, previously uncharted, strong-gravity regime. 

Utilizing this data to constrain new physics beyond GR requires obtaining alternative predictions for the leading-order corrections in the above gravitational EFT. As large-curvature regimes reveal the nonlinear character of (beyond) GR dynamics, such predictions require numerical relativity simulations, cf.~\cite{Shibata:1995we,Baumgarte:1998te,Pretorius2005,Pretorius:2005gq} for pioneering work in numerical GR. 
It is, therefore, crucial to find a well-posed numerical evolution for said leading-order EFT corrections. The existence of a well-posed initial value problem (IVP), see~\cite{Sarbach2012,Isenberg2014} for reviews in GR, could pose a restriction for any viable theory. Certainly, a well-posedness IVP (for physically meaningful sets of initial data) is necessary to perform any stable numerical evolution.
\\

First numerical simulations have been achieved in specific beyond-GR theories such as dynamical Chern-Simons~\cite{Okounkova2019,Okounkova2020}, Einstein-dilaton-Gauss-Bonnet~\cite{Ripley2020a,Ripley2020b,Okounkova:2020rqw,East2021,Silva:2020omi}, Horndeski theories \cite{Figueras:2020dzx}, K-essence \cite{Bezares:2020wkn}, and a study of EFT terms at 
quartic order in curvature~\cite{Cayuso:2020lca}.
A stable numerical evolution is guaranteed by either iterative treatment \cite{Okounkova2019,Okounkova2020,Okounkova:2020rqw}, a dampening of high-frequency modes \cite{Cayuso:2020lca}, or by an established well-posed evolution at weak coupling~\cite{Kovacs2019, Kovacs2020a, Kovacs2020b,East2021}. In contrast, at strong coupling the onset of ill-posed regimes has been observed in~\cite{Ripley2020a,Ripley2020b}.

Here, we investigate Quadratic Gravity (QG), i.e., the gravitational EFT including all independent terms up to quadratic order in curvature -- sometimes also referred to as Stelle-gravity~\cite{Stelle:1976gc, Stelle1978}. At the formal level, it has been shown -- without any constraint to weak coupling -- that QG admits a well-posed IVP~\cite{Noakes:1983}, see also~\cite{Morales:2018imi}.
\\

Accounting for EFT corrections to GR naturally implies higher-order equations of motion.
The latter are theoretically disfavored by the Ostrogradski theorem~\cite{Ostragradski1850} which states that non-degenerate higher-derivative theories result in linearized degrees of freedom with opposite-sign kinetic terms. Any non-vanishing coupling between these modes implies the onset of a linear instability, cf.~eg.~\cite{Becker:2017tcx,Anselmi:2018ibi,Donoghue:2019fcb,Salvio:2019ewf} for related recent developments in the context of a unitary quantum evolution of QG.

We emphasize that an Ostrogradski (in)stability and well-posedness are not necessarily related.
On the one hand, the Ostrogradski theorem is a physical statement: theories that exhibit opposite-sign kinetic terms can develop physical instabilities. On the other hand, well-posedness is a mathematical property of partial-differential equations (PDEs), crucial for numerical simulations, but not necessarily related to physical implications. In particular, the same theory can admit, both, well-posed and ill-posed IVPs. We add that the absence of \emph{any} well-posed IVP would also signal a true physical shortcoming.

In fact, the present work can be viewed as a proof of principle for a higher-derivative gravitational theory with Ostrogradski ghosts which nevertheless admits a well-posed IVP and in which we can thus simulate spacetime dynamics numerically. The proof in \cite{Noakes:1983} establishes the existence of a well-posed IVP at the level of the four-dimensional equations of motion. As in GR, it remains non-trivial to translate it into (3+1) form to obtain a well-posed IVP suitable for actual numerical evolution. 
Here, we do so in the reduced sector of spherically-symmetric dynamics. This allows us to present the first nonlinear evolution of physical initial data in QG.
\\

The rest of the paper is organized as follows: in Section~\ref{sec:qg_noakes}, we review the well-posed IVP formulation of QG, cf.~\cite{Noakes:1983}; in Section~\ref{sec:cartoon_ss}, we use the Cartoon method~\cite{Alcubierre1999, Pretorius2005} to reduce the evolution equations to spherical symmetry and perform the order-reduction in the symmetry-reduced setup; in Section~\ref{sec:result}, we present the resulting \emph{stable} numerical evolution for perturbations of flat spacetime as well as for perturbations of the Schwarzschild solution and find no indications for ill-posed behavior; in Section~\ref{sec:conclusion} we conclude with a summary and discuss the implications of our results for future works.

As for conventions, we use the $(-,+,+,+)$ signature, we work in geometrized units where ($c=1$, $G=1$), and
use Latin letters as Lorentzian spacetime indices.
Round (square) brackets denote (anti-)symmetrization of the enclosed pair of indices.

\section{Quadratic Gravity and the Noakes equations}
\label{sec:qg_noakes}

Quadratic Gravity (QG) incorporates the leading-order, i.e., curvature-squared, corrections to GR and can be parameterized by the action
\begin{align}
    S_\text{QG} = \int d^4x \sqrt{|g|}
    \left[
        \frac{1}{16\pi G}R
        +\alpha R_{ab}R^{ab}
        - \beta R^2
    \right]\;.
\end{align}
Note that we have chosen the minus sign in the last term to agree with conventions in the equations of motion in~\cite{Noakes:1983}.
The first term is the common Einstein-Hilbert term where we have, for the present section, re-instated Newton's constant $G$. In 4D, the most general corrections of quadratic order in curvature can be parameterized by the two dimensionless constants $\alpha$ and $\beta$. A potential Riemann-squared term can be rewritten into the former through the Gauss-Bonnet identity.

The associated equations of motion for quadratic gravity (QG) contain (up to) 4th-order derivative terms~\cite{Stelle:1976gc} and read
\begin{align}
\label{eq:eom}
    2\,T_{ab}=
    &(\alpha - 2\beta)\nabla_a\nabla_b R
    -\alpha\Box R_{ab}
    -\left(\frac{1}{2}\alpha - 2\beta\right)g_{ab}\Box R
    \notag\\
    &+2\alpha R^{cd}R_{acbd}
    -2\beta RR_{ab}
    -\frac{1}{2}(\alpha R^{cd}R_{cd} - \beta R^2)
    \notag\\
    &+\frac{1}{16\pi G}\left(R_{ab}-\frac{1}{2}g_{ab}R\right)\;.
\end{align}
Here, we have included an energy-momentum tensor $T_{ab}$ for potential matter sources.

In the linearized theory, it has been shown, cf.~\cite{Stelle:1976gc}, that, in addition to the massless spin-2 mode of GR, the linearized Ricci scalar and linearized traceless Ricci tensor propagate a massive scalar and massive spin-2 modes with respective masses
\begin{align}
\label{eq:masses}
    m_0^2 = \frac{1}{32\pi G(3\beta - \alpha)}\;,
    \quad\quad\quad
    m_2^2 = \frac{1}{16\pi G\alpha}\;.
\end{align}
The massive spin-2 mode is an Ostrogradski ghost, i.e., in the linearized theory, its kinetic term has the opposite sign in comparison to the massless graviton.
In~\cite{Noakes:1983}, Noakes finds that the nonlinear evolution can be formulated with the same degrees of freedom. Following this insight, the Ricci scalar $\mathcal{R}\equiv g^{ab}R_{ab}$ and the traceless Ricci tensor $\widetilde{\mathcal{R}}_{ab}\equiv R_{ab} - 1/4\,g_{ab}\,R$ can be elevated to independent variables, as indicated by the curly notation. From here on, these should no longer be evaluated on the metric. Rather, they are treated as independent evolution variables. Re-expressing the parameters $\alpha$ and $\beta$ by the two masses $m_0^2$ and $m_2^2$, the equations of motion can be separated into a trace (2nd equation below) and a traceless (3rd equation below) part, i.e.,
\begin{align}
\label{eq:metric}
    R_{ab}(g) =&\;\widetilde{\mathcal{R}}_{ab} + \frac{1}{4} g_{ab}\mathcal{R}\;,
    \\[0.5em]
\label{eq:trace}
    \Box\,\mathcal{R}=&\;m_0^2\mathcal{R} 
    + 2T^c_{\phantom{c}c}\;,
    \\[0.5em]
\label{eq:traceless}
    \Box\,\widetilde{\mathcal{R}}_{ab}=&\;m_2^2\widetilde{\mathcal{R}}_{ab}
    + 2T^\text{(TL)}_{ab}
    \notag\\
    &- \frac{1}{3}\left(\frac{m_2^2}{m_0^2}-1\right)\left(\nabla_a\nabla_b \mathcal{R} - \frac{1}{4}g_{ab}m_0^2 \mathcal{R}\right)
    \notag\\
    &+ 2\widetilde{\mathcal{R}}^{cd}C_{acbd}
    -\frac{1}{3}\left(\frac{m_2^2}{m_0^2}+1\right)\mathcal{R}\widetilde{\mathcal{R}}_{ab}
    \notag\\
    &-2\widetilde{\mathcal{R}}_{a}^{\phantom{a}c}\widetilde{\mathcal{R}}_{bc}
    +w\frac{1}{2}g_{ab}\widetilde{\mathcal{R}}^{cd}\widetilde{\mathcal{R}}_{cd}\;.
\end{align}
Here, we have supplemented the trace and traceless equations by the definition of the Ricci curvature in terms of the metric (on the left-hand side) and in terms of the fiducial variables (on the right-hand side), cf. Eq.~\eqref{eq:metric}. The latter provides a 2nd-order evolution equation for the metric in which the fiducial variables appear as sources, i.e., without derivatives. The matter sources are also split into trace and traceless part (indicated by a $^{(TL)}$ superscript) and, in turn, only source the fiducial variables.
Furthermore, we have introduced the Weyl-tensor $C_{abcd}$ for brevity. The latter can equivalently be expressed in terms of $R_{abcd}$, $\widetilde{\mathcal{R}}_{ab}$, and $\mathcal{R}$ as
\begin{align}
    C_{abcd} = R_{abcd} 
    + g_{b[c}\widetilde{\mathcal{R}}_{a]d}
    + g_{d[a}\widetilde{\mathcal{R}}_{c]b}
    +\frac{1}{6}g_{a[d}g_{b]c}\mathcal{R}\;.
\end{align}
Therefore, the metric appears in the trace and traceless equation, cf.~Eqs.~\eqref{eq:trace}-\eqref{eq:traceless}, only as part of the covariant derivatives as well as in $R_{abcd}$.

The trace and traceless Ricci variables $\mathcal{R}$ and $\widetilde{\mathcal{R}}_{ab}$ appear merely as `fiducial sources', i.e., as terms without derivatives. Thus, the metric equation, cf.~Eq.~\eqref{eq:metric}, can be treated as in GR where (generalized) harmonic coordinates allow to express the Ricci tensor as a strongly hyperbolic quasi-linear 2nd-order differential operator. More explicitly, the Ricci tensor on the left-hand side of Eq.~\eqref{eq:metric} can be expressed as
\begin{align}
\label{eq:Ricci-harmonic}
    R_{ab}(g) = 
    -\frac{1}{2}g^{cd}g_{ab,cd} 
    + g_{c(a}F^c_{\phantom{c},b)} + \mathcal{O}^{1}_{ab}(\partial g)\;,
\end{align}
where we use the usual comma-notation for partial derivatives and $F^a = -g^{cd}\Gamma^a_{cd}$ is a gauge potential which, provided one works in (generalized) harmonic gauge, i.e., $F^a=0$ ($F^a=\text{const}$), reveals the strongly hyperbolic quasi-linear character.
Additional lower-order derivative terms are denoted by $\mathcal{O}^1_{ab}(\partial g)$.

With (generalized) harmonic coordinates at hand and by expanding the Riemann tensor as well as all covariant derivatives in terms of the metric, the evolution of the variables $g_{ab}$, $\mathcal{R}$, and $\widetilde{\mathcal{R}}_{ab}$ can be written as
\begin{align}
    \label{eq:metric-1}
    g^{cd}g_{ab,cd} &= 
    -2\widetilde{\mathcal{R}}_{ab} 
    -\frac{1}{2} g_{ab}\mathcal{R} + \mathcal{O}^1_{ab}(\partial g)\;,
    \\[0.5em]
    \label{eq:trace-1}
    g^{cd}\mathcal{R}_{,cd} &= m_0^2\mathcal{R}\;,
    \\[0.5em]
    \label{eq:traceless-1}
    g^{cd}\widetilde{\mathcal{R}}_{ab,cd} &=  
    \mathcal{O}^2_{ab}(\partial\partial\mathcal{R},\partial\widetilde{\mathcal{R}},\partial\partial g)\;.
\end{align}
Again, we summarize lower-order terms with $\mathcal{O}^1_{ab}(\partial g)$ and $\mathcal{O}^2_{ab}(\partial\partial\mathcal{R},\partial\widetilde{\mathcal{R}},\partial\partial g)$ with the notation in brackets indicating the highest order of derivatives of these terms. As will become clear below, their explicit form is irrelevant regarding well-posedness. Nevertheless, we explicitly provide their form in App.~\ref{app:lower-order-terms-explicit}, where we also correct some typos in comparison to~\cite{Noakes:1983,Morales:2018imi}.

The above system is not yet of \emph{diagonal} quasi-linear form due to both $\partial\partial\mathcal{R}$- and $\partial\partial g$-terms appearing in the traceless equation, cf.~Eq.~\eqref{eq:traceless-1}. However, the system is amenable to diagonalization because of the lack of $\partial\widetilde{\mathcal{R}}$-contributions in the (already diagonal-form) equations for $\mathcal{R}$ (Eq.~\eqref{eq:trace-1}) and $g$ (Eq.~\eqref{eq:metric-1}), see~\cite{Noakes:1983} for more formal and general statements. More explicitly, the given PDEs can be diagonalized by introducing extra variables $V_a \equiv \mathcal{R}_{,a}$ and $h_{abc}\equiv g_{ab,c}$ and adding derivatives of the former two equations to the system, i.e.,
\begin{align}
    g^{mn}V_{a,mn} &= 
    \mathcal{O}_{a}(\partial V,h)\;,
    \\
    g^{mn}h_{abc,mn} &=
    \mathcal{O}_{abc}(\partial h)\;,
    \\
    g^{mn}\widetilde{\mathcal{R}}_{ab,mn} &=  
    \mathcal{O}^2_{ab}(\partial V,\partial h,\partial\widetilde{\mathcal{R}})\;.
\end{align}
This extended system is now of diagonal quasi-linear form and the standard theorems for hyperbolicity~\cite{leray.pdf.1953} apply, cf.~\cite{Noakes:1983} for a more detailed proof and our separate publication~\cite{Lim2021future} for a full (3+1)-formulation of QG.
\\

Treating $\widetilde{\mathcal{R}}_{ab}$ and $\mathcal{R}$ as independent variables, Eqs.~\eqref{eq:metric}-\eqref{eq:traceless} only describe the physical evolution of QG if additional constraints guarantee that  $\widetilde{\mathcal{R}}_{ab}$ and $\mathcal{R}$ equate to their metric counterparts. This can be captured by introducing a constraint variable $\mathcal{C}_{ab} = G_{ab} - \widetilde{\mathcal{R}}_{ab} + 1/4\,g_{ab}\mathcal{R}$, describing the deviation of the fiducial variables from the physical Einstein tensor. Demanding that $\mathcal{C}_{ab}$ and its first time derivative vanish (Bianchi constraints), ensures that the initial data is physical. Similarly, since the above formulation as a quasi-linear system requires harmonic coordinates, we need to ensure that the initial data does so too. Overall, the initial-data constraints read
\begin{align}
\label{eq:harmonic-constraint-1}
    F^a &= 0\;,
    \\
\label{eq:harmonic-constraint-2}
    \pounds_n F^a &= 0\;,
    \\
\label{eq:bianchi-constraint-1}
    \mathcal{C}_{a\phantom{b};b}^{\phantom{a}b} &=0\;,
    \\
\label{eq:bianchi-constraint-2}
    (\pounds_n\mathcal{C}\indices{_a^b})_{;b} &=0\;,
\end{align}
where $\pounds_n$ denotes Lie derivatives along a timelike normal vector $n^a$, orthogonal to the initial-data hypersurface. A similar order-reduction as for the evolution equations above, cf.~\cite{Noakes:1983}, ensures that also the propagation of the constraints can be written as a quasilinear diagonal 2nd-order system.

In addition, initial data has to obey the usual Gauss-Codazzi (or Hamiltonian and shift) constraints of GR which read
\begin{align}
\label{eq:hamiltonian-constraint}
    n^a n^b\mathcal{C}_{ab} &= 0\;,
    \\
\label{eq:shift-constraint}
    n^a h_c^b \mathcal{C}_{ab} &= 0\;.
\end{align}
Finally, also the physical constraints are preserved in the temporal direction (secondary constraints).
\\

Counting the number of pieces of independent initial data reveals that the nonlinear theory retains the same number of degrees of freedom as the linearized one. Eqs.~\eqref{eq:metric}-\eqref{eq:traceless}t constitute a set of 20 2nd-order evolution equations (equivalent to the 10 4th-order equations in Eq.~\eqref{eq:eom}). At the same time, the harmonic constraints (Eqs.~\eqref{eq:harmonic-constraint-1}-\eqref{eq:harmonic-constraint-2}), the Bianchi constraints (Eqs.~\eqref{eq:bianchi-constraint-1}-\eqref{eq:bianchi-constraint-2}), and the Gauss-Codazzi constraints (Eqs.~\eqref{eq:hamiltonian-constraint}-\eqref{eq:shift-constraint}) require $8$ relations among the initial data (including the respective secondary constraints), each. Overall this amounts to $40-24=16$ pieces of independent initial data, corresponding to eight degrees of freedom, i.e., a massless spin-2 (two degrees of freedom), a massive scalar (one degree of freedom), and a massive spin-2 mode (five degrees of freedom). We will keep track of this counting in the order reduction and the reduction to spherical symmetry below.

\section{Reduction to spherical symmetry via the Cartoon method}
\label{sec:cartoon_ss}

Unfortunately, choosing coordinates in which spherical symmetry is explicit makes it impossible to maintain the harmonic gauge condition $\Box  x^a=0$~\cite{Sorkin:2009bc}. However, as we have seen in the last section, the latter is a crucial to achieve a well-posed formulation, cf. discussion below Eq.~\eqref{eq:Ricci-harmonic}. We, therefore, remain in Cartesian coordinates $(t,x,y,z)$ and follow the Cartoon method~\cite{Alcubierre1999, Pretorius2005} to make use of the Killing vector fields associated with spherical symmetry, i.e.,
\begin{align*} 
    \xi_1^\mu &= x(\pmb{\partial}_y)^\mu - y(\pmb{\partial}_x)^\mu\;,
    \\
    \xi_1^\mu &= y(\pmb{\partial}_z)^\mu - z(\pmb{\partial}_y)^\mu\;,
    \\
    \xi_1^\mu &= z(\pmb{\partial}_x)^\mu - x(\pmb{\partial}_z)^\mu\;.
\end{align*}
Note that the bold font $\pmb{\partial}_{x,y,z}$ indicate the basis one-forms and \emph{not} partial derivatives. Expanding the respective vanishing Lie-derivatives $\mathcal{L}_{\xi_i}\pmb{X}=0$ acting on tensorial objects $\pmb{X}$, one can re-express partial derivatives in two of the spatial directions, for instance $(y,z)$, in terms of the third, for instance $x$. For scalars $\Phi$, vectors $\Psi_a$ and tensors $\Pi_{ab}$ which are preserved along the Killing vector field, one finds
\begin{align*}
    \partial_{y}\Phi &= \frac{y}{x}\partial_x\Phi\;,
    \\
    \partial_{y}\Psi_a &= \frac{1}{x}\Big(
        y\,\partial_x\Psi_a + \Psi_x\,\delta_a^y - \Psi_y\,\delta_c^x
    \Big)\;,
    \\
    \partial_{y}\Pi_{ab} &= \frac{1}{x}\Big(
        y\,\partial_x \Pi_{ab} - 2x\,\delta^x_{(a}\Pi_{b)y} + 2y\,\delta^y_{(a}\Pi_{b)x}
    \Big)\;,
\end{align*}
and equivalently for $(y\leftrightarrow z)$. These relations allow us to reduce all spatial derivatives in Eqs.~\eqref{eq:metric-1}-\eqref{eq:traceless-1}
to those with respect to a single coordinate, e.g., $x$. 

Furthermore, a coordinate transformation from a coordinate system in which the spherical symmetry is manifest, i.e., transforming $\overline{X} = (t,r,\theta,\phi)$ back to Cartesian coordinates $X=(t,x,y,z)$, gives symmetry relations between tensor components via
\begin{align}
\label{eq:cartoon-relations}
    \Pi_{\overline{a}\overline{b}} = 
    \frac{\partial X^a}{\partial \overline{X}^{\overline{a}}}
    \frac{\partial X^b}{\partial \overline{X}^{\overline{b}}}
    \Pi_{ab}\;.
\end{align}
For the present case of spherical symmetry, transformation of the symmetry identities $\Pi_{t\theta} = 0$, $\Pi_{t\phi} = 0$, $\Pi_{r\theta} = 0$, $\Pi_{r\phi}=0$, $\Pi_{\theta\phi}=0$, and $\Pi_{\theta\theta}\sin^2\theta = \Pi_{\phi\phi}$ back to Cartesian coordinates implies the relations
\begin{align}
    \Pi_{ty} &= \frac{y\,\Pi_{tx}}{x}\;,
    \quad\quad\quad\quad\;\,
    \Pi_{tz} = \frac{z\,\Pi_{tx}}{x}\;,
    \\
    \Pi_{xy} &= \frac{xy\left(\Pi_{xx}-\Pi_{yy}\right)}{x^2-y^2}\;, \;
    \Pi_{xz} = \frac{xz\left(\Pi_{xx}-\Pi_{yy}\right)}{x^2-y^2}\;,
    \\
    \Pi_{yz} &= \frac{yz\left(\Pi_{xx}-\Pi_{yy}\right)}{x^2-y^2}\;,
    \\
    \Pi_{zz} &= \frac{(x^2 - z^2)\Pi_{yy}-(y^2 - z^2)\Pi_{xx}}{x^2-y^2}\;.
\end{align}
Naturally, all of the above also holds for raised indices. For $\widetilde{\mathcal{R}}_{ab}$, we can additionally make use of tracelessness to remove one further component.

With the above relations, the evolution equations in Eqs.~\eqref{eq:metric-1}-\eqref{eq:traceless-1} (with the explicit form of lower-order terms provided in App.~\ref{app:lower-order-terms-explicit}) can be expanded into evolution equations for only eight independent variables which we group into
\begin{align}
    \mathbf{u} = (\mathcal{R},g_{tt},g_{tx},g_{xx},g_{yy})
    \;\text{and}\;
    \mathbf{v} = (\widetilde{\mathcal{R}}_{tt},\widetilde{\mathcal{R}}_{tx},\widetilde{\mathcal{R}}_{xx}),
\end{align}
according to whether their associated evolution equations are already quasi-linear or not.
The set of eight 2nd-order equations takes the form
\begin{align}
     \partial_t^2\mathbf{u}&= \mathcal{O}\left(\mathbf{u},\,\mathbf{v},\,\partial_t{\mathbf{u}}\right)\;,
    \\
    \partial_t^2\mathbf{v}&= \mathcal{O}\left(\mathbf{u},\,\mathbf{v},\,\partial_t{\mathbf{u}},\,\partial_t{\mathbf{v}},\,\partial_t^2{\mathbf{u}}\right)\;.
\end{align}
Unfortunately, the explicit expressions are too large to meaningfully be displayed here. Instead, we provide them, along with all the subsequent reductions, in ancillary files~\footnote{See the \texttt{GitHub} repository (\url{https://github.com/aaron-hd/QG-sphSymm-ancillary}) for the \texttt{Mathematica} \cite{Mathematica} script and ancillary files. Parts of the derivation make use of the \texttt{xAct} package \cite{Martin-Garcia:2007bqa} (\url{http://www.xact.es/}).} and restrict the following discussion to a schematic form. We note that, in distinction to Sec.~\ref{sec:qg_noakes}, the above notation only keeps track of the order of time derivatives such that all instances of $\mathcal{O}$ can potentially contain up to 2nd-order spatial derivatives.

The above 2nd-order evolution system naively propagates 16 free initial data functions, i.e., 8 degrees of freedom. These are subject to \emph{physical constraints}, which we will come back to in Sec.~\ref{sec:physicalConstraints}. For now, we only want to keep track of additional \emph{auxiliary constraints} which appear due to the order-reduction below.

\subsection{Reduction to quasi-linear 2nd-order form}

In analogy to the 4D-diagonalization procedure in~\cite{Noakes:1983}, we introduce additional auxiliary variables
\begin{align}
\dot{\mathbf{u}}=(\dot{\mathcal{R}},\,\dot{g}_{tt},\,\dot{g}_{tx},\,\dot{g}_{xx},\,\dot{g}_{yy}) \equiv \partial_t\mathbf{u}
\end{align}
and differentiate the first two equations by time.
Adding the resulting equations to the evolution system results in
\begin{align}
    \partial_t^2\dot{\mathbf{u}}&= \mathcal{O}\left(\mathbf{u},\,\mathbf{v},\,\dot{\mathbf{u}},\,\partial_t\dot{\mathbf{u}},\,\partial_t\mathbf{v}\right)\;,
    \\
    \partial_t^2\mathbf{v}&= \mathcal{O}\left(\mathbf{u},\,\mathbf{v},\,\dot{\mathbf{u}},\,\partial_t\dot{\mathbf{u}},\,\partial_t\mathbf{v}\right)\;,
    \\
    \partial_t\mathbf{u} &\equiv \dot{\mathbf{u}}\;,
    \\
    \partial_t\dot{\mathbf{u}}&= \mathcal{O}\left(\mathbf{u},\,\mathbf{v},\,\dot{\mathbf{u}}\right)\;.
\end{align}
In the above, we have added the definitions $\dot{\mathbf{u}} \equiv \partial_t\mathbf{u}$ to the evolution equations. Indeed, these defining equations for the auxiliary variables act as 1st-order evolution equations for $\mathbf{u}$ while the original evolution equations become 1st-order \emph{auxiliary constraints} on $\dot{\mathbf{u}}$ since there are other 2nd-order evolution equations for the latter. Thereby, the evolution system remains consistent with $21-5 = 16$ free functions of initial data.

\subsection{Reduction to 1st-order form}
By introducing eight further auxiliary variables, 
\begin{align}
\ddot{\mathbf{u}}\equiv \partial_t\dot{\mathbf{u}}
\quad\text{and}\quad
\dot{\mathbf{v}}\equiv \partial_t \mathbf{v}\;,
\end{align}
the set of evolution equations is cast into a from in which it becomes purely 1st-order in time, i.e.,
\begin{align}
    \label{eq:1stOrder-start}
    \partial_t\ddot{\mathbf{u}}&= \mathcal{O}\left(\mathbf{u},\,\mathbf{v},\,\dot{\mathbf{u}},\,\ddot{\mathbf{u}},\,\dot{\mathbf{v}}\right)\;,
    \\
    \partial_t\dot{\mathbf{v}}&= \mathcal{O}\left(\mathbf{u},\,\mathbf{v},\,\dot{\mathbf{u}},\,\ddot{\mathbf{u}},\,\dot{\mathbf{v}}\right)\;,
    \\
    \partial_t\dot{\mathbf{u}} &\equiv \ddot{\mathbf{u}}\;,
    \\
    \partial_t\mathbf{u} &\equiv \dot{\mathbf{u}}\;,
    \\
    \partial_t\mathbf{v} &\equiv \dot{\mathbf{v}}\;,
    \label{eq:1stOrder-end}
    \\
    \label{eq:auxiliary-constraints}
    \ddot{\mathbf{u}}&= \mathcal{O}\left(\mathbf{u},\,\mathbf{v},\,\dot{\mathbf{u}}\right)\;.
\end{align}
These evolution equations, which we have supplemented by the defining equations for the auxiliary variables $\dot{\mathbf{u}}$, $\dot{\mathbf{v}}$, and $\ddot{\mathbf{u}}$, now denote $21$ 1st-order evolution equations.

It is now apparent that the original evolution equations for $\mathbf{u}$ take the role of five \emph{auxiliary constraints}, cf.~Eq.~\eqref{eq:auxiliary-constraints}, such that the overall number of free functions to be specified as initial data remains $21-5 = 16$. 

We emphasize, once more, that we have hidden up-to-2nd-order spatial derivatives in the notation such that the auxiliary constraint in Eq.~\eqref{eq:auxiliary-constraints} is by no means algebraic but rather a 2nd-order spatial ODE.
\\

A formal proof of well-posedness of the above evolution equations would either require a further reduction of the remaining higher-order spatial derivatives, or to analyse strong hyperbolicity as a first-order in time and arbitrary-order in space (FTNS) system along the lines of \cite{Gundlach:2005ta,Richter:2013naa,Hilditch:2014naa}. Instead of pursuing such a formal proof, we focus on demonstrating a stable numerical evolution in Sec.~\ref{sec:result}.
Here, and in the following, we use the term `numerical stability' to distinguish our setup from physical stability and emphasize that a reliable numerical exploration of physical stability requires a numerically stable setup.
To be explicit, by numerical stability, we refer to the apparent absence of growth modes in the constraints.

\subsection{Physical Constraints}
\label{sec:physicalConstraints}
In spherical symmetry, the 24 physical constraints reduce to 12 since in each four-component constraint equation two Cartoon relations (arising from a relation equivalent to Eq.~\eqref{eq:cartoon-relations} but for vectors) can be used. We are therefore left with $21-5(\text{auxiliary})-12(\text{physical}) = 4$ independent pieces of initial data, i.e., two degrees of freedom. 

More explicitly, the Hamiltonian and shift constraint reduce to
\begin{align}
    \mathcal{C}_{tt}&\equiv
    G_{tt} 
    - \widetilde{\mathcal{R}}_{tt} 
    + \frac{1}{4}\,g_{tt}\mathcal{R} = 0 \label{eq:hamiltonioanConstraint_sphSymm}\;,
    \\
    \mathcal{C}_{tx}&\equiv
    G_{tx}
    - \widetilde{\mathcal{R}}_{tx} 
    + \frac{1}{4}\,g_{tx}\mathcal{R}\label{eqn:shift_const}\;.
\end{align}
These physical constraints reproduce (some components of) the auxiliary constraints, cf., Eq.~\eqref{eq:auxiliary-constraints}, upon using $R(g) = \mathcal{R}$. Since $R(g) = \mathcal{R}$ and $\widetilde{R}_{ab}(g) = \widetilde{\mathcal{R}}_{ab}$ these are, of course, the very relations which are supposed to be enforced by the auxiliary constraints.
This implies that Hamiltonian and momentum constraints are automatically
fulfilled once the auxiliary constraints are fulfilled. 
\\

In the subsequent numerical analysis, we will monitor the Hamiltonian constraint to confirm the absence of any growth modes. Indeed, we find no indication for ill-posed behavior.

Solving the constraints, both physical and auxiliary, is a non-trivial task and will be addressed in future work. Below, we will focus on perturbations of exact solutions of the theory for which all constraints are fulfilled. Moreover, we focus on vacuum solutions of GR which, since they are Ricci-flat, are exact solutions of QG as well.

\section{Numerically stable evolution}
\label{sec:result}

Having derived a set of 1st-order (in time) evolution equations for the spherically-symmetric sector, cf.~Eqs.~\eqref{eq:1stOrder-start}-\eqref{eq:1stOrder-end}, we now proceed to solve these numerically.
We evolve random-noise perturbations of (i) flat spacetime, and (ii) Schwarzschild spacetime, cf.~Sec.~\ref{sec:method} for details of the setup. 
In Sec.~\ref{sec:conv}, we perform convergence tests to ensure that our system satisfies the expected order of convergence.
In Sec.~\ref{sec:HamCon}, we monitor the Hamiltonian constraint $\mathcal{C}_{tt}$, cf.~Eq.~\eqref{eq:hamiltonioanConstraint_sphSymm}, in order to confirm the absence of any growth modes.

\subsection{Numerical method}
\label{sec:method}

We use a fourth-order finite difference method to evaluate spatial derivatives and a fourth-order Runge-Kutta method to evolve in time. The computational domain is chosen as $x\in (0,10M]$ ($x\in (0,10]$ for flat spacetime), working in units of the mass $M$. 
We evolve all our equations in a unigrid with $N_x = 1025$ points. Hence the grid resolution is $\Delta x \simeq 0.01\,M$ with the Courant–Friedrichs–Lewy condition~\cite{Courant:1967} set to 0.25. Therefore, as we increase $N_x$ (or decrease $\Delta x$), the time discretization $\Delta t$ decreases. 
\\

We perform several simulations 
in order to test whether the numerical evolution 
of random initial data close to (i) flat spacetime and (ii) Schwarzschild spacetime
is consistent with a well-posed IVP.
For flat spacetime, we initialize the evolution at
\begin{align}
    \mathbf{u}_{0} &= (\mathcal{R},\,g_{tt},\,g_{tx},\,g_{xx},\,g_{yy}) = (0,-1,0,1,1)\;,
    \notag\\
    \mathbf{v}_{0} &= \dot{\mathbf{v}}_{0} = \dot{\mathbf{u}}_{0} = \ddot{\mathbf{u}}_{0} = 0\;.
    \label{eq:init-flat}
\end{align}
For Schwarzschild spacetime, we work in
in Cartesian Kerr-Schild coordinate such that
\begin{align}
    g_{tt,0} &= \sqrt{\frac{r}{r+2M}}\;,
    \quad
    g_{tx,0} = \frac{2M}{r} \frac{x}{r+2M}\;,
    \notag\\
    g_{xx,0} &= 1+ \frac{2Mx^2}{r^3}\;,
    \quad
    g_{yy,0} = 1+ \frac{2My^2}{r^3}\;,
    \notag\\[0.5em]
    \mathcal{R}_{0} &=0\;, 
    \notag\\[0.5em]
    \mathbf{v}_{0} &= \dot{\mathbf{v}}_{0} = \dot{\mathbf{u}}_{0} = \ddot{\mathbf{u}}_{0} = 0\;,
    \label{eq:init-BH}
\end{align}
where $r=\sqrt{x^2+y^2+z^2}$ and $M$ is the mass of the Schwarzschild black hole.

Given the respective background  $(\mathbf{u}_0,\mathbf{v}_0,\dot{\mathbf{u}}_0,\dot{\mathbf{v}}_0,\ddot{\mathbf{u}}_0)$, cf.~Eq.~\eqref{eq:init-flat} for flat and Eq.~\eqref{eq:init-BH} for Schwarzschild spacetime, we add random noise to all components of initial data, i.e,
\begin{align}
    \mathbf{u}^{i}(t=0) &= \mathbf{u}^{i}_0 + A_{\textrm{noise}} \textrm{RAND}(x)\;\;\forall\; i=1,\dots,5\;,
    \notag\\
    \mathbf{v}^{i}(t=0) &= \mathbf{v}^{i}_0 + A_{\textrm{noise}} \textrm{RAND}(x)\;\;\forall\; i=1,\dots,3\;,
    \notag \\
    \dot{\mathbf{u}}^{i}(t=0) &= \dot{\mathbf{u}}^{i}_0 + A_{\textrm{noise}} \textrm{RAND}(x)\;\;\forall\; i=1,\dots,5\;,
    \notag\\
    \dot{\mathbf{v}}^{i}(t=0) &= \dot{\mathbf{v}}^{i}_0 + A_{\textrm{noise}} \textrm{RAND}(x)\;\;\forall\; i=1,\dots,3\;,
    \notag\\
    \ddot{\mathbf{u}}^{i}(t=0) &= \ddot{\mathbf{u}}^{i}_0 + A_{\textrm{noise}} \textrm{RAND}(x)\;\;\forall\; i=1,\dots,5\;.
    \label{eq:noise}
\end{align}
Here, $A_{\textrm{noise}}$ is a noise amplitude which we vary from $10^{-10}$ to $10^{-5}$ and $\textrm{RAND}(x)$ generates random values between -1 and 1. 
In Sec.~\ref{sec:conv}, we present the respective
self-convergence tests to validate our numerical implementation and to verify the expected rate of convergence with decreased noise amplitude. 
Since the random noise violates the constraints, the above simulations constitute a robust stability test.
In Sec.~\ref{sec:HamCon}, we verify explicitly that the Hamiltonian constraint does not exhibit any growth modes that would signal an ill-posed IVP.

\subsection{Convergence Tests}
\label{sec:conv}

\begin{figure}[!t]
    \centering
    \includegraphics[trim={0.5cm 0cm 2.5cm 1cm},clip,width=\columnwidth]{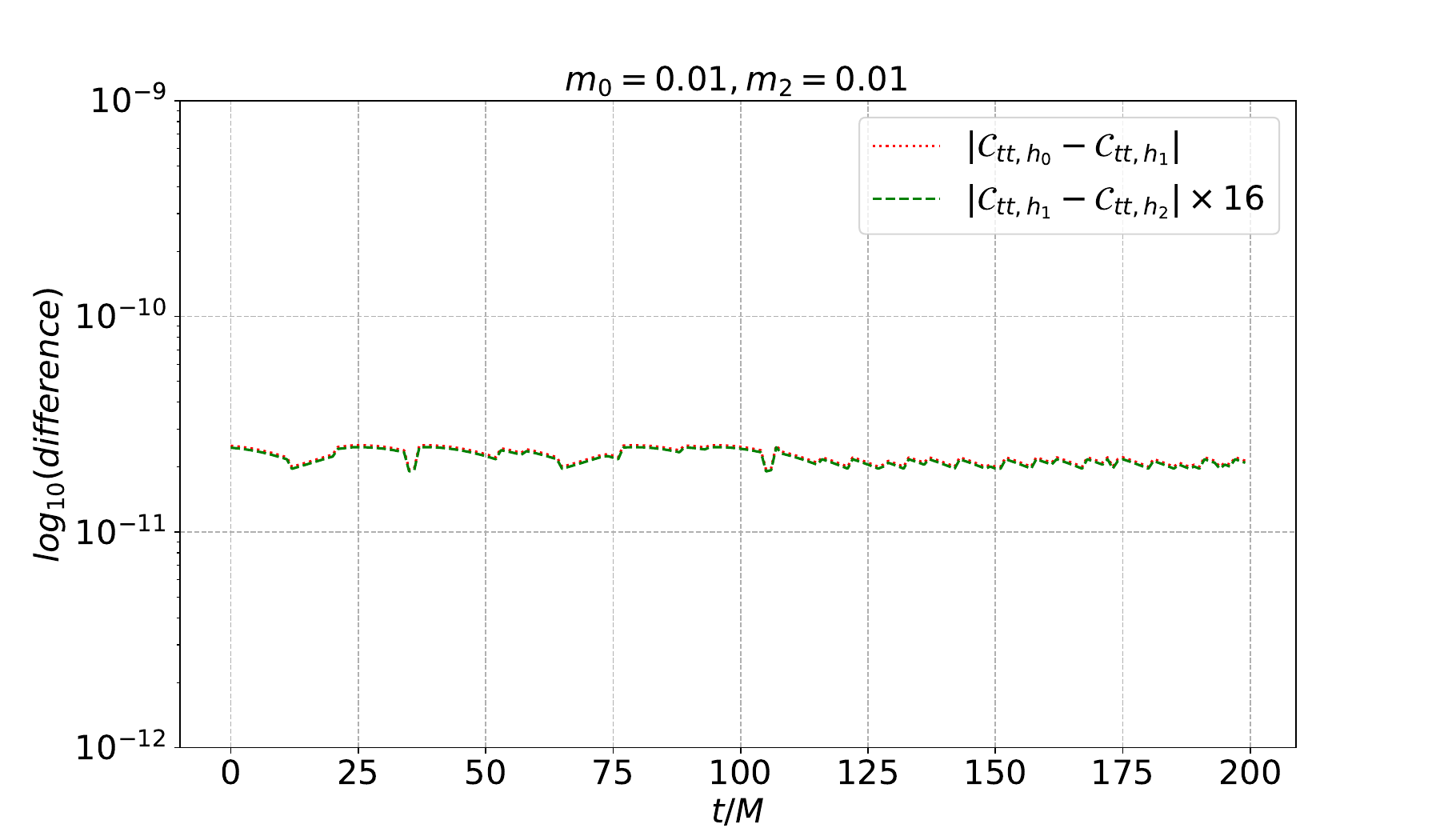}
\caption{
\label{fig:convergence}
   Self-convergence test for Schwarzschild spacetime, evolved until $t/M \simeq 200$ with $M$ the mass of the black hole. We choose $m_0 = 0.01, m_2 = 0.01$ in units of $M$. We plot the differences of $\mathcal{C}_{tt}$ with different resolutions. The differences remain small during the entire evolution. The smaller resolution difference is rescaled by a factor of 16, in agreement with the expected convergence rate of our implementation.
}
\end{figure}

We perform standard convergence tests to confirm the validity of our implementation and to demonstrate consistency with convergence of numerical errors to a well-posed continuum system.
Since we apply fourth-order finite-difference  stencils, the expected convergence rate is four.
We choose a coarsest resolution of $h_0 = 0.01$ and then decrease to different resolutions $h_i$ with $i = 0,1,\dots,5$ such that $h_{i+1} = h_i/2$. Further, we decrease the grid spacing by a factor of 2 when we increase the resolution.

Standard convergence tests have to be performed with respect to a specified norm, suitable for the given system of evolution equations.
If second-order spatial derivatives of the evolution variables dominate the numerical evolution, the standard $L_2$ norm may not be appropriate and a different norm, such as $H_1$, may be more suitable, cf.~\cite{Calabrese2006,Babiuc2008} for more comprehensive discussion.
Indeed, the implemented system of evolution equations, cf.~Eqs.~\eqref{eq:1stOrder-start}-\eqref{eq:1stOrder-end}, may contain second-order spatial derivatives. Due to the complexity of the system, it is non-trivial to identify which of the evolution variables are dominated by such second-order spatial derivatives. 
Therefore, we perform and compare convergence tests with respect to both the conventional $L_2$ as well as the $H_1$ norm.

Fig.~\ref{fig:convergence} summarizes the result of these convergence test.
We plot differences (in $L_2$ norm) of the constraint value, $\mathcal{C}_{tt}$, with different resolutions as a function of time. The smaller resolution difference is re-scaled by a factor of $16$, which corresponds to the expected fourth-order convergence rate, as detailed below. 
The differences remain small and both lines lie on top of each other which confirms the fourth-order convergence.
\\

\begin{figure}[!t]
    \centering
    \includegraphics[trim={0.5cm 0cm 1.5cm 0.8cm},clip,width=\columnwidth]{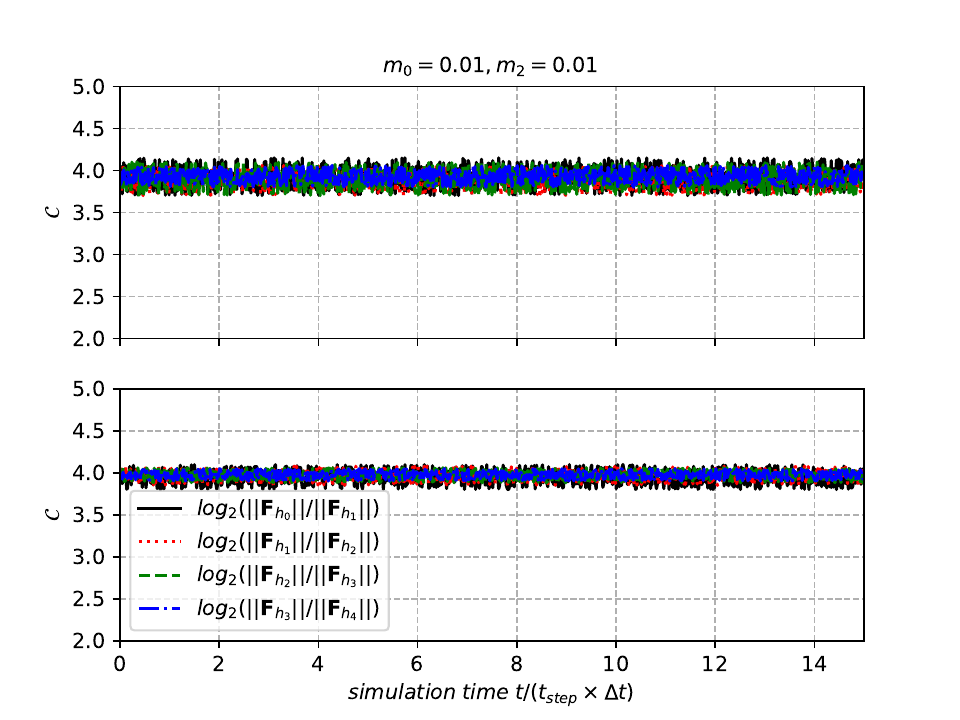}
\caption{
\label{fig:ExactConvRatio}
   Exact convergence test. Both $L_2$ (Above) and $H_1$ (Below) are computed and show expected convergence ratio.
}
\end{figure}
\begin{figure}[!t]
    \centering
    \includegraphics[trim={0.5cm 0cm 1.5cm 0.8cm},clip,width=\columnwidth]{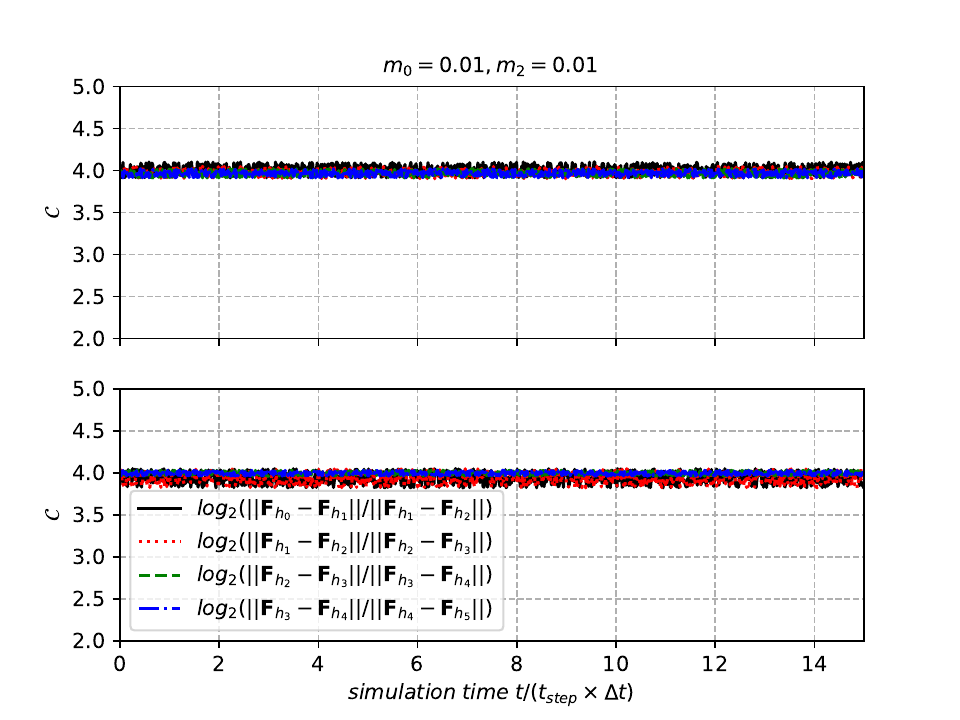}
\caption{
\label{fig:SelfConvRatio_flat}
   Self-convergence test with flat spacetime as a function of simulation time. We compute both $L_2$ (upper panel) and $H_1$ (lower panel) norms. Both cases exhibit the expected convergence ratio.
}
\end{figure}

To be specific,
the self-convergence ratio is given by
\begin{align}
    \label{eqn:SelfConv_ratio}
    \mathcal{C}_{\textrm{self}} = \log_2 \frac{||\mathbf{F}_{h_i} - \mathbf{F}_{h_{i+1}}||_q}{||\mathbf{F}_{h_{i+1}} - \mathbf{F}_{h_{i+2}}||_q} ,
\end{align}
where $\mathbf{F}$ is the state vector for all evolution variables, i.e., $\mathbf{F} = (\mathbf{u}, \mathbf{v},  \dot{\mathbf{u}},\dot{\mathbf{v}}, \ddot{\mathbf{u}} )^T$, and $|| \cdot ||_q$ is a general expression for different norms. In the following, we denote with $|| \cdot ||_{L_2}$ and $|| \cdot ||_{H_1}$ the $L_2$ and $H_1$ norm, respectively. These norms are computed in a discrete approximation that replaces each continuum norm~\cite{Giannakopoulos2020}.
\\
The exact convergence ratio, with $\mathbf{F}_\textrm{exact} = 0$, is given by
\begin{align}
    \label{eqn:ExactConv_ratio}
    \mathcal{C}_{\textrm{exact}} = 
    \log_2 \frac{||\mathbf{F}_{h_i} - \mathbf{F}_\textrm{exact}||_q}{||\mathbf{F}_{h_{i+1}} - \mathbf{F}_\textrm{exact}||_q} = 
    \log_2 \frac{||\mathbf{F}_{h_i} ||_q}{||\mathbf{F}_{h_{i+1}} ||_q} \; .
\end{align}
Determining $\mathcal{C}_{\textrm{exact}}$ only requires two different resolutions and is thus numerically cheaper than determining $\mathcal{C}_\textrm{self}$. 
Given the employed fourth-order scheme, the expected convergence rate is four, in both cases.
\\

\begin{figure}[!t]
    \centering
    \includegraphics[trim={0.5cm 0cm 1.0cm 0.8cm},clip,width=\columnwidth]{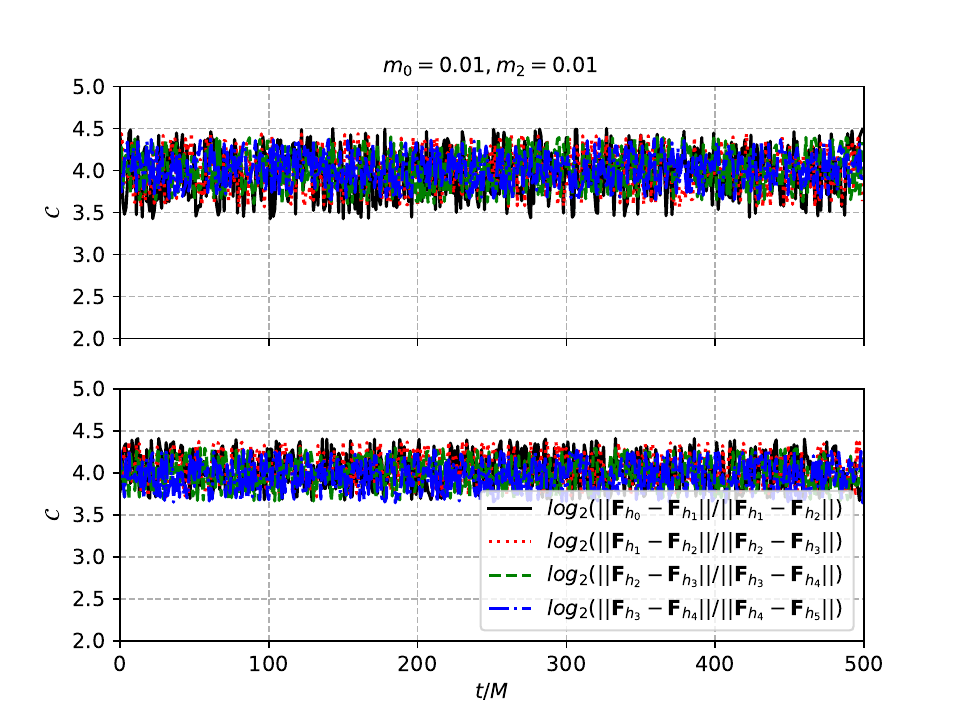}
\caption{
\label{fig:SelfConvRatio_SchwBH}
   Self-convergence test for Schwarzschild spacetime as a function of physical time. We compute both $L_2$ (upper panel) and $H_1$ (lower panel) norms. Both cases exhibit the expected convergence ratio.
}
\end{figure}
The appropriate rescaling of the random noise with decreased resolution, is determined by the respective norm. 
Let $A_{h_i}$ be an amplitude of the random noise associated with the respective resolution $h_i$.
\\
For the $L_2$ norm, we have
\begin{align}
\label{eqn:rescaling_l2}
    \mathcal{C}_{\textrm{exact}} = \log_2 \frac{||\mathbf{F}_{h_i} ||_{L_2}}{||\mathbf{F}_{h_{i+1}} ||_{L_2}} \sim \log_2 \frac{\mathcal{O}(A_{h_i})}{\mathcal{O}(A_{h_{i+1}})}.
\end{align}
Hence, for the fourth-order numerical scheme at hand, we need to multiply the amplitude of random noise with a
factor of $1/16$ when doubling the resolution. 
\\
For the $H_1$ norm, we have
\begin{align}
    \label{eqn:ExactConv_ratio_H1}
    \mathcal{C}_{\textrm{exact}} = &
    \log_2 \frac{||\mathbf{F}_{h_i} ||_{H_1}}{||\mathbf{F}_{h_{i+1}} ||_{H_1}}
    \notag\\
     = & \log_2 \frac{||\mathbf{F}_{h_i} ||_{L_2} + ||\nabla \mathbf{F}_{h_i} ||_{L_2}}{||\mathbf{F}_{h_{i+1}} ||_{L_2}+||\nabla\mathbf{F}_{h_{i+1}} ||_{L_2}}
    \notag\\
    \sim &\log_2 \frac{\mathcal{O}(A_{h_i})}{2 \mathcal{O}(A_{h_{i+1}})}\;,
\end{align}
where $\nabla \mathbf{F}$ is spatial derivative of $\mathbf{F}$ (not a covariant derivative). To compute $\nabla \mathbf{F}$, the centered, second-order finite-difference method is applied. In this case, the norm is dominated by the derivative term.
Hence, for the fourth-order numerical scheme at hand, we need to multiply the amplitude of random noise with a
factor of $1/32$ when doubling the resolution. 
\\
In both cases, the arguments also hold for the self-convergence ratio. 
\\

For all convergence tests, we choose a coarsest noise amplitude of $A_{h_0}=10^{-5}$. The self-convergence tests, are performed for flat and Schwarzschild spacetime in the $L_2$ and $H_1$ norm, cf.~Figs.~\ref{fig:SelfConvRatio_flat} and~\ref{fig:SelfConvRatio_SchwBH}. The exact convergence tests for the $L_2$ and $H_1$ norm are shown in Fig.~\ref{fig:ExactConvRatio}. 
In all cases, we verify the expected convergence ratio, both in $L_2$ and $H_1$, although the $L_2$ result is slightly more noisy. This is consistent with a well-posed evolution with respect to both norms and we refrain from conclusively determining whether second-order spatial derivatives dominate the evolution. 
To do so, we would need to examine the system of evolution equations, cf.~Eqs.~\eqref{eq:1stOrder-start}-\eqref{eq:1stOrder-end}, term by term.

\subsection{Absence of growth modes in the constraint violations}
\label{sec:HamCon}
\begin{figure}
    \centering
    \includegraphics[trim={0.6cm 0cm 2.9cm 1cm},clip,width=\linewidth]{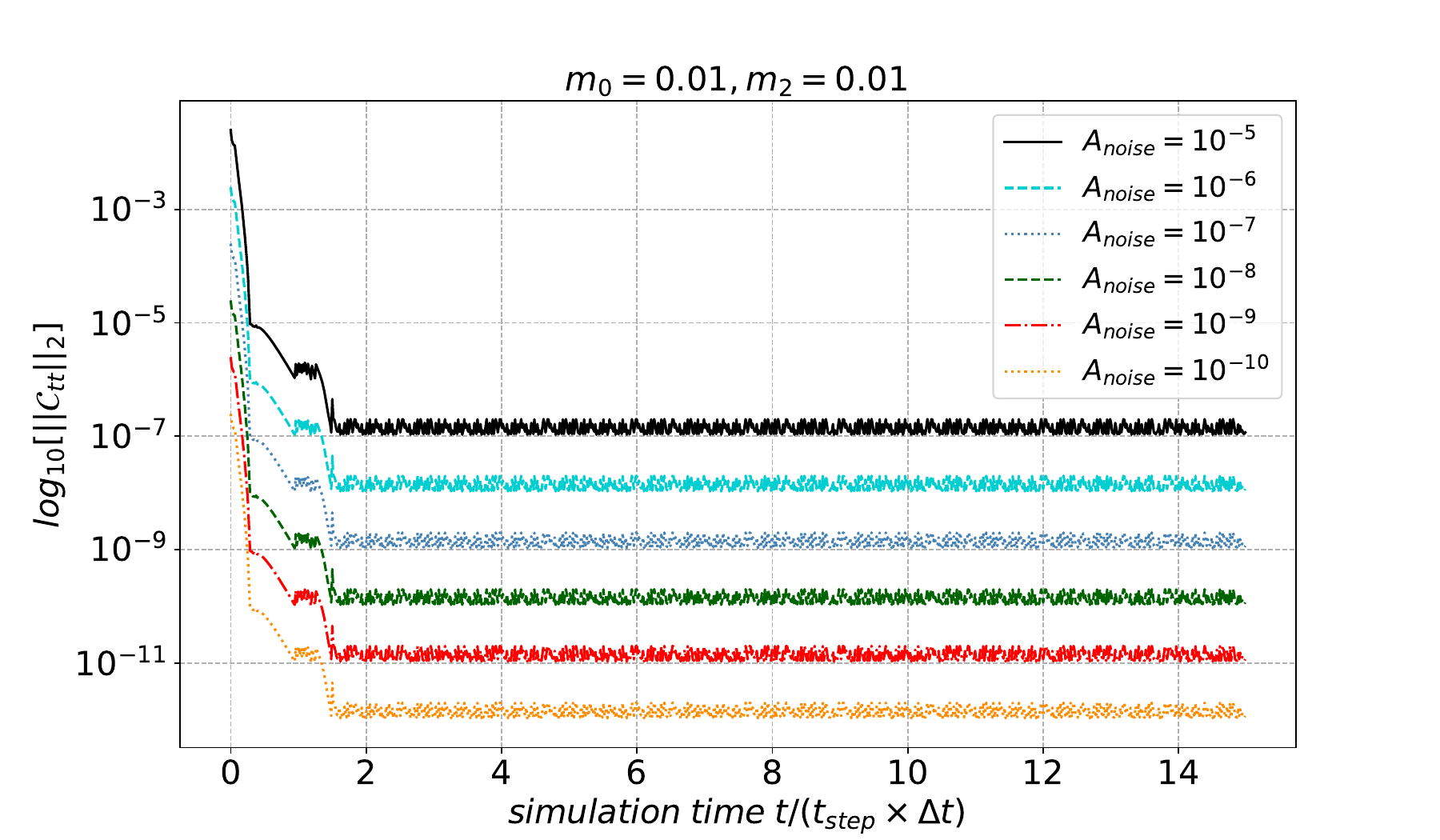}
\caption{
\label{fig:noise}
    Constraint plot ($L_2$-norm of the Hamiltonian constraint in Eq.~\eqref{eq:hamiltonioanConstraint_sphSymm}) for flat-space initial data, performing a noise test with different noise amplitudes, ranging from $10^{-5}$ to $10^{-10}$, top to bottom. Each curve represents an increase by a factor of ten in the initial amplitude over the curve below.
}
\end{figure}

Here, we monitor the behavior of the Hamiltonian constraint, cf.~Eq.~\eqref{eq:hamiltonioanConstraint_sphSymm}, for sufficiently long evolution time. 
Since the other constraints (and the evolution equations) are coupled, it can be expected that violations of any other constraint will percolate into the Hamiltonian constraint.
The absence of growth modes in the constraint violations suggests that we are evolving a well-posed IVP.
\\

For flat spacetime, we evolve initial data up to simulation time,  $t / (t_{step} \times \Delta t) \simeq 15$. We find that the Hamiltonian constraint first decays and then stabilizes, cf.~Fig.~\ref{fig:noise}. This indicates that the evolution time is sufficiently long for the constraint violations to settle into a near-stable state. There appear to be no growth modes in the constraint for flat-space initial data, at least during the monitored time.
We conclude that the performed noise test finds no indications of numerically unstable or ill-posed behavior. 
\\
\begin{figure}
    \centering
    \includegraphics[trim={0.9cm 0cm 2.9cm 1cm},clip,width=\columnwidth]{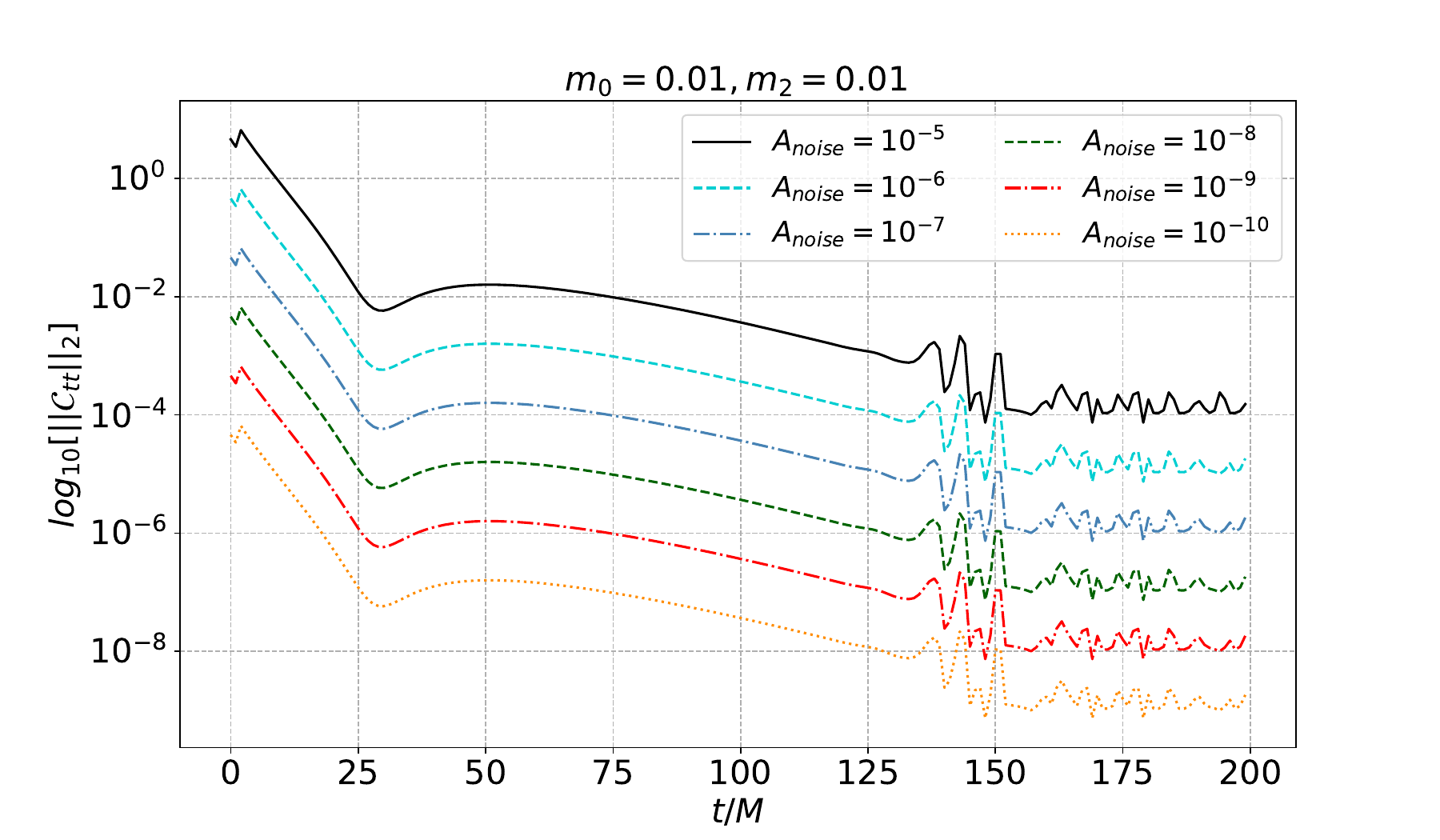}
\caption{
\label{fig:schwBH_noise}
    Constraint plot ($L_2$-norm of the Hamiltonian constraint in Eq.~\eqref{eq:hamiltonioanConstraint_sphSymm}) for Schwarzschild initial data with added noise. Different noise amplitudes were chosen as described in Fig.~\ref{fig:noise}
}
\end{figure}

Similar results persist for initial data corresponding to a Schwarzschild black hole with mass $M$.
We evolve simulations until $t/M \simeq 200$ which suffices for the constraint violations to first decay and the to settle into a near-stable state. We find no indication of numerically unstable or ill-posed behaviour 

\begin{figure}
    \centering
    \includegraphics[trim={0.9cm 0cm 2.9cm 1cm},clip,width=\columnwidth]{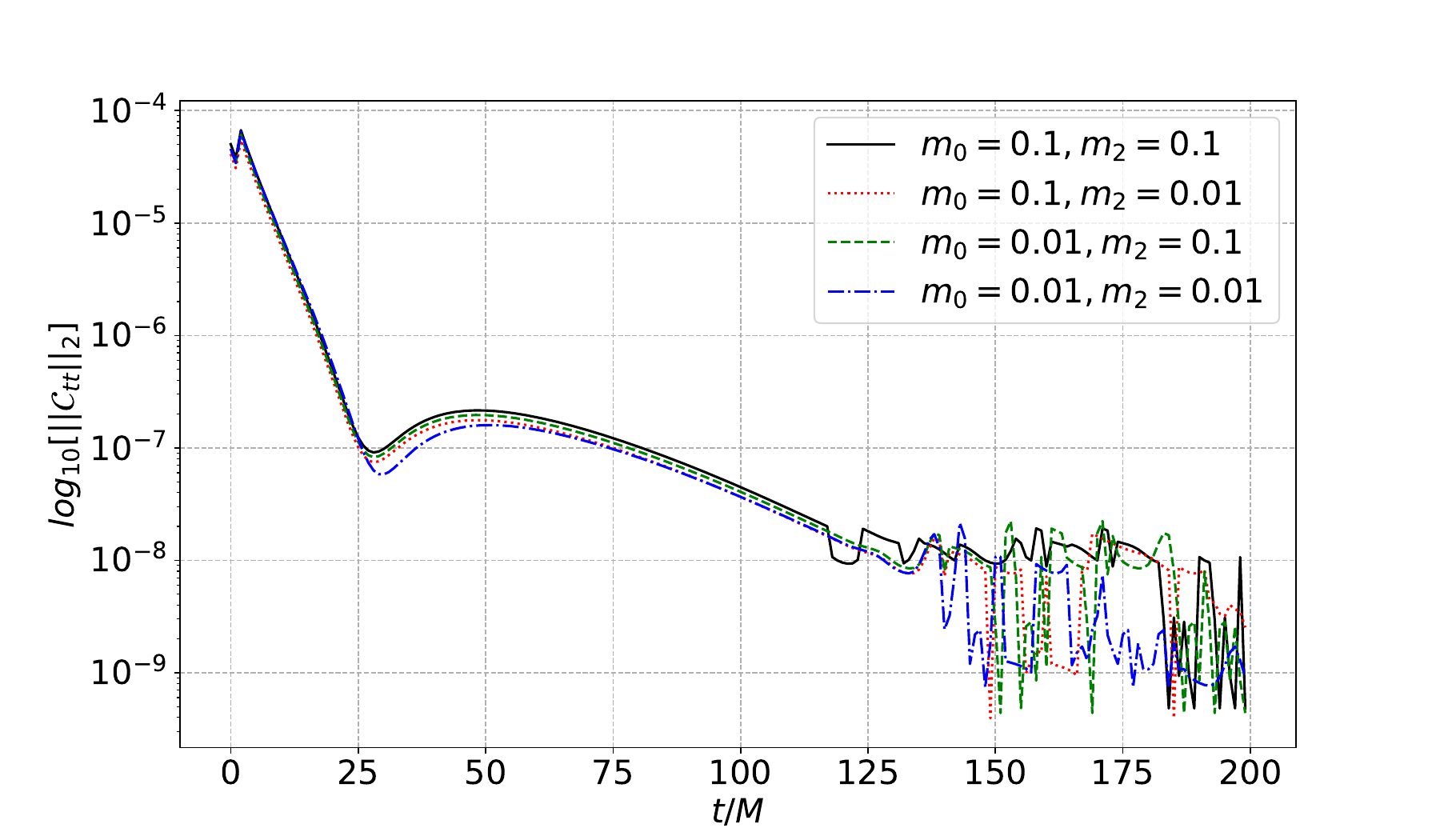}
\caption{
\label{fig:schwBH}
    Constraint plot for Schwarzschild initial data for different QG mass parameters $m_0$ and $m_2$ (in units of the Schwarzschild mass $M$).
}
\end{figure}

In addition, we vary the masses $m_0$ and $m_2$ of the QG spin-0 and spin-2 modes, cf.~Eq.~\eqref{eq:masses}, in order to confirm that the absence of growth modes in the constraint evolution persists for a range of values for $m_0$ and $m_2$ in the vicinity of the Schwarzschild mass $M$, cf.~Fig.~\ref{fig:schwBH}.
We observe \emph{numerically} stable evolution for all tested values of $m_0$ and $m_2$. The constraint decays and stabilizes over the investigated time.

We emphasize that this does not test physical stability. Even in mass-ranges (for the QG masses $m_0$ and $m_2$) in which the Schwarzschild solution could become \emph{physically} unstable, we expect the constraint evolution to be \emph{numerically} stable. In particular, even during a potential physical decay of Schwarzschild spacetime, possibly to some other solution of QG~\cite{Pravda2017,Podolsky:2019gro}, the constraints should be preserved. We will investigate such physical (in)stability in future work.
\\

Overall, we conclude that, both for flat-space and for Schwarzschild initial data, no indications of numerically unstable behavior are found. In conjunction with the convergence tests performed in Sec.~\ref{sec:conv}, this strongly suggests that the evolution is indeed well-posed.

\section{Conclusion}
\label{sec:conclusion}

As a proof of principle for well-posed numerical evolution in beyond-GR theories with higher-curvature operators, we successfully obtain the first fully nonlinear time evolution in Quadratic Gravity, i.e., for a gravitational theory including the lowest-order EFT corrections to GR. The system of evolution equations is obtained by (i) use of harmonic gauge to treat the Ricci scalar and traceless Ricci tensor as independent variables, cf.~\cite{Noakes:1983}, (ii) reduction to spherical symmetry, which preserves harmonic gauge by use of the Cartoon method~\cite{Alcubierre1999, Pretorius2005}, and (iii) order-reduction to a set of manifestly 1st-order (in time) evolution equations.

We perform non-linear numerical simulations that are fully consistent with an underlying well-posed IVP for  physically significant initial data.
In particular, we observe numerically stable dynamics, i.e., the absence of any growth modes, in the spherically-symmetric sector for random perturbations about flat spacetime and about the Schwarzschild solution. (Being Ricci-flat, the latter is also a solution of QG.)
\\

This opens up several opportunities for future work. 
As a direct application, we will investigate physically (un)stable branches of Schwarzschild BHs and other exotic BHs in QG, cf.~\cite{Lu:2015cqa,Lu:2015psa,Lu:2017kzi,Holdom:2016nek,Kokkotas:2017zwt,Bonanno:2019rsq,Podolsky:2019gro}, to determine the final state of spherical gravitational collapse. 

As a long-term goal, the present study strongly motivates that one can also establish a well-posed IVP in full 3+1 dimensions, within computational infrastructures such as~\cite{Fernando2019,einsteintoolkit}. 
We emphasize that while the proof in \cite{Noakes:1983} guarantees the existence of a well-posed IVP also in (3+1) dimensions, the explicit construction of such a formulation remains non-trivial.
Establishing such a well-posed IVP will eventually enable us to perform binary-black-hole mergers to extract GW signals in this theory.

Finally, the presented methodology is, in principle, applicable also to other theories with higher-order equations of motion. As long as the order-reduced equations of motion are amenable to diagonalization to quasi-linear form, other higher-derivative theories may also admit a well-posed IVP.
\\

\paragraph*{Acknowledgements.}
We are grateful for hospitality at the Perimeter Institute for Theoretical Physics, where the original idea for this manuscript was developed. Research at Perimeter Institute is supported in part by the Government of Canada through the Department of Innovation, Science and Economic Development Canada and by the Province of Ontario through the Ministry of Colleges and Universities.
We thank Frans Pretorius and Justin Ripley for correspondence regarding the Cartoon method as well as Eric W. Hirschmann and Astrid Eichhorn for discussions.
HL is supported by the LANL ASC Program and LDRD grants 20190021DR. This work used resources provided by the LANL Institutional Computing Program. LANL is operated by Triad National Security, LLC, for the National Nuclear Security Administration of the U.S.DOE  (Contract No. 89233218CNA000001). AH is supported by a Royal Society Newton International fellowship [NIF/R1/191008]. This work is authorized for unlimited release under LA-UR-21-22739

\appendix

\section{Explicit form of lower-order terms}
\label{app:lower-order-terms-explicit}

The explicit form of the lower-order terms in the metric Eq.~\eqref{eq:metric-1} is given by
\begin{align}
    \mathcal{O}^1_{ab}(\partial g)
    =&\;
    F^c\Gamma_{(ab)c}
    +2g^{ed}\Gamma^c_{e(a}\Gamma_{b)cd}
    +g^{cd}\Gamma^e_{ad}\Gamma_{ecb}\;,
\end{align}
where the first term vanishes upon use of harmonic coordinates.
This comes about from the well-known expansion of the Ricci-tensor in harmonic coordinates.

The explicit form of $\mathcal{O}^2_{ab}(\partial\mathcal{R},\partial\widetilde{\mathcal{R}},\partial\partial g)$ in the traceless equation, cf.~Eq.~\eqref{eq:traceless-1}, is given by
\begin{widetext}
\begin{align}
    \mathcal{O}^2_{ab}(\partial\partial\mathcal{R},\partial\widetilde{\mathcal{R}},\partial\partial g) 
    =&
    - \frac{1}{3}\left(\frac{m_2^2}{m_0^2} - 1\right)\mathcal{R}_{,ab}
    \underline{
    - \widetilde{\mathcal{R}}_{cd}g^{cf}g^{ed}\left(
        g_{ef,ab} + g_{ab,ef}
        - g_{ae,fb} - g_{fb,ae}
    \right)
    }
    \notag\\&
    \dashuline{
    +\frac{1}{6}\left(\widetilde{\mathcal{R}}_{ab}g^{ef}g^{cd}g_{ef,cd}\right)
    +g^{ef}g^{cd}\widetilde{\mathcal{R}}_{e(a}\left(
        g_{df,b)c}
        +g_{b)f,cd}
        -g_{b)d,cf}
        \right)
    }
    \notag\\&
    \underline{
    + 2\widetilde{\mathcal{R}}_{cd}\,g^{ce}\Big[
        2\Gamma^d_{f[e}\,\Gamma^f_{a]b}
        + g^{df}\,g^{mn}g_{fn,[a}\left(g_{e]m,b}+g_{bm,e]}-g_{e]b,m}\right)
    \Big]
    }
    \notag\\&
    \dashuline{
    +g^{cd}\Big[
        2\Gamma^n_{c(a}\left(
            \Gamma^m_{dn}\widetilde{\mathcal{R}}_{b)m}
            +\widetilde{\mathcal{R}}_{b)n,d}
        \right)
        +2\Gamma^m_{d(b}\left(
            \Gamma^n_{a)c}\widetilde{\mathcal{R}}_{mn}
            -\widetilde{\mathcal{R}}_{a)m,c}
        \right)
    }
    \notag\\&\quad\quad\quad
    \dashuline{
        +g^{ef}g^{mn}g_{fn,c}\widetilde{\mathcal{R}}_{e(b}\left(
            g_{dm,a)}+g_{a)m,d}-g_{a)d,m}
        \right)
    \Big]
    }
    \notag\\&
    +m_2^2\widetilde{\mathcal{R}}_{ab}
    + 2T^\text{(TL)}_{ab}
    + \frac{1}{12}\left(\frac{m_2^2}{m_0^2}-1\right)g_{ab}m_0^2 \mathcal{R}
    \notag\\
    &-\frac{1}{3}\left(\frac{m_2^2}{m_0^2}+1\right)\mathcal{R}\widetilde{\mathcal{R}}_{ab}
    -2g^{cd}\widetilde{\mathcal{R}}_{ac}\widetilde{\mathcal{R}}_{bd}
    +\frac{1}{2}g_{ab}\widetilde{\mathcal{R}}^{cd}\widetilde{\mathcal{R}}_{cd}\;.
\end{align}
\end{widetext}
We have sorted the contributions in terms of their order in time derivatives: the first two lines collect all 2nd-order terms; lines three to five all 1st-order terms; the last two lines collect the 0th-order contributions.

Furthermore, we have underlined terms arising from ${\widetilde{\mathcal{R}}^{cd}R_{abcd}}$ (continuous) and from ${\Box\widetilde{\mathcal{R}}_{ab}}$ (dashed). The former agree with~\cite{Noakes:1983} (apart from minor sign typos) but the latter do not. In any case, as discussed in the main text, a modification of these terms (as long as their derivative order is preserved) does not impact Noakes' proof of well-posedness.
\\

\bibliography{References}

\end{document}